\documentclass[conference]{IEEEtran}
\IEEEoverridecommandlockouts
\usepackage{latexsym}
\usepackage{graphicx}
\usepackage{amsfonts,amssymb,amsmath}
\def\BibTeX{{\rm B\kern-.05em{\sc i\kern-.025em b}\kern-.08em T\kern-.1667em\lower.7ex\hbox{E}\kern-.125emX}}

\usepackage{algorithmic}
\usepackage{algorithm}
\usepackage{subfig}
\usepackage{siunitx}
\usepackage{cite}

\usepackage[acronym]{glossaries}
\newcommand{\newac}{\newacronym}
\newcommand{\ac}{\gls}

\makeglossaries

\newac{los}{\text{LoS}}{line-of-sight}
\newac{nlos}{\text{NLoS}}{non-line-of-sight}

\DeclareMathAlphabet{\mathbit}{OML}{cmr}{bx}{it}
\DeclareMathAlphabet{\mathsf}{OT1}{cmss}{m}{n}
\DeclareMathAlphabet{\mathTXf}{OT1}{cmss}{bx}{it}

\DeclareMathOperator{\Transpose}{T}

\DeclareMathOperator*{\argmin}{arg\,min}

\DeclareMathAlphabet{\mathpzc}{OT1}{pzc}{m}{it}

\newcommand{\Tr}{{\Transpose}}

\usepackage{anyfontsize}

\begin{document}

\title{UAV-aided Wireless Node Localization Using Hybrid Radio Channel Models
}

\author{
\IEEEauthorblockN{Omid Esrafilian, Rajeev Gangula, and David Gesbert}
\IEEEauthorblockA{Communication Systems Department, EURECOM, Sophia Antipolis, France\\
\text{\{esrafili, gangula, gesbert\}@eurecom.fr}
}
}

\maketitle

\begin{abstract} 
This paper considers the problem of ground user localization based on received signal strength (RSS) measurements obtained by an unmanned aerial vehicle (UAV). We treat UAV-user link channel model parameters and antenna radiation pattern of the UAV as unknowns that need to be estimated. A \textit{hybrid} channel model is proposed that consists of a traditional path loss model combined with a neural network approximating the UAV antenna gain function. With this model and a set of offline RSS measurements, the unknown parameters are estimated. We then employ the particle swarm optimization (PSO) technique which utilizes the learned hybrid channel model along with a 3D map of the environment to accurately localize the ground users. The performance of the developed algorithm is evaluated through simulations and also real-world experiments.

\end{abstract}





\section{Introduction}\label{sec:Intro}

In a wireless localization system, nodes with perfectly-known positions known as anchor nodes (which can be stationary or mobile) collect various radio measurements from the emitted radio frequency (RF) signals from the users in the network, and use them for localization purposes. Various measurements  such  as  received signal strength (RSS), time-of-arrival (TOA), angle of arrival (AOA),  etc.,  can  be  obtained from the RF signals by the anchor nodes \cite{zekavat2011handbook,delRauLop}.


On the other hand, advancement in robotic technologies and miniaturization of wireless equipment have made it possible to have  flying radio networks (FRANs), where wireless connectivity  to ground  users can be provided by aerial base stations (BSs) that  are  mounted  on unmanned  aerial  vehicles  (UAVs) \cite{GanEsaGes,MozSaadBennNamDebb}.
Advantage of FRANs include, fast  and  dynamic  network  deployment  during  an  emergency or temporary crowded events, providing connectivity in areas lacking network infrastructure, etc. While in terrestrial radio access networks static BSs are used as anchor nodes, in FRANs UAV BSs can be used as mobile anchor nodes.

Localization of ground users using RSS measurements collected by 
aerial UAV anchor nodes
has  gained  interest  recently \cite{Ref5_sallouha2017aerial,Ref19_lima2019support,Ref9_artemenko2015evaluation,Ref10_ji2019fair,Ref6_koohifar2018autonomous,ShahidSoltan,EsrGanGesAsil,esrafilian20203d}.
The main advantage of using UAV BS anchors in localization compared to static BSs is that UAV BSs with their inherent 3D mobility can collect 
radio measurements in difference geographic locations which improves the localization performance. Generally, RSS measurements are easy to obtain in many wireless networks and does not require stringent synchronization and calibration constraints associated with timing based measurements. The works in \cite{Ref10_ji2019fair,Ref6_koohifar2018autonomous,ShahidSoltan}
assumed that the UAV flies high enough so that the RSS of the 
UAV-user or Air-to-ground (A2G) link is modeled as a simple \ac{los} channel. This LoS assumption is generally not valid in urban scenarios as A2G links are often blocked by city buildings. To over come this, the authors of \cite{EsrGanGesAsil,esrafilian20203d} have used a segmented pathloss model that differentiates between LoS and \ac{nlos} channel conditions and showed improvement in localization performance. Moreover, in \cite{esrafilian20203d} it is shown that by exploiting the 3D city map which contains the building locations and height information, one can significantly improve the localization performance. 

One common assumption in all the works in  \cite{Ref5_sallouha2017aerial,Ref19_lima2019support,Ref9_artemenko2015evaluation,Ref10_ji2019fair,Ref6_koohifar2018autonomous,ShahidSoltan,EsrGanGesAsil,esrafilian20203d}, is that the UAV is assumed to have a perfect isotropic radiating antennas. However, in reality this is not true and several complications arise with UAV BSs as opposed to static BSs: 
a) The UAV altitude and heading changes depending on its mobility pattern, hence the antenna gain changes with the UAV location and orientation \cite{7228721}
b) The radiation pattern of the antenna mounted on UAV is affected by the chassis, and hence it is difficult to measure the actual antenna pattern while UAV is flying \cite{7510362,8402597}. 
The work in \cite{CheDevSinGuv} has demonstrated that the 3D radiation pattern of an
antenna mounted on a drone can significantly influence the RSS of the A2G link. Therefore, it is important to consider this in a practical localization system using UAV BS anchors.

The impact of the antenna
radiation pattern in A2G channels in a 3D localization system using  time-difference-of-arrival (TDOA) measurements has been studied in 
\cite {SinYapISm}. However, the authors assume that the radiation pattern of the UAV antenna is known. To the best of our knowledge, localizing users with a UAV with an unknown radiation pattern and RSS measurements has not been studied before. Specifically, our contributions are as follows:
\begin{itemize}
\item  The unknown antenna radiation pattern of the UAV during the flight is parameterized by a neural network by feeding RSS measurements in the training phase.
\item An optimization framework is proposed capable of using the trained model, which characterizes the path loss and the antenna gain pattern, along with the 3D map of the environment to improve the localization performance.
\end{itemize}

\begin{figure}[t]
\begin{centering}
\includegraphics[width=0.8\columnwidth]{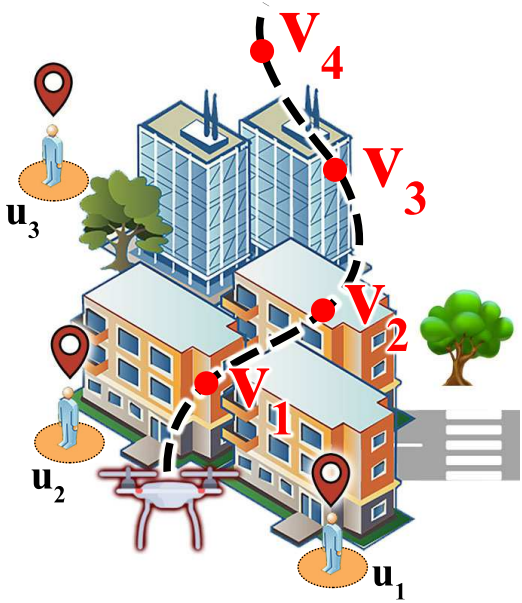}
\par\end{centering}
\caption{UAV-aided ground user localization system.\label{fig:SystemModel}}
\end{figure}

\section{System Model and Problem Formulation}\label{sec:SysModel}
We consider a scenario similar to the one illustrated in Fig. \ref{fig:SystemModel}, where a UAV BS that is
connected to $K$ ground level users in an urban area
consisting of a number of city buildings. The users are
spread over the city and ${\bf u}_{k}=[x_{k},y_{k}]^{\Tr}\in\mathbb{R}^{2},\,k\in[1,K]$
denotes the $k$-th user's location. The users are considered static and their locations are unknown.
A 3D map of the environment where the UAV and users are located is assumed to be available.

The aim of the UAV is to estimate
the unknown user locations based on RSS measurements taken in $N$ different time steps from the users. In the $n$-th time step, the UAV/drone position is denoted by 
${\bf{v}}_n=[x[n],y[n],z[n]]^{\Tr}\in\mathbb{R}^{3}$.
We assume that the UAV is equipped with a GPS receiver, hence
${\bf{v}}_n,\forall n$ is known.



\subsection{Channel Model}
We now describe the radio channel model between the UAV and ground users. Note that the channel parameters and the UAV antenna pattern are unknown and need to be learned. In general, the channel between UAV position ${\bf{v}}$ and user location ${\bf{u}}$ in dB can be modeled as
\begin{equation}
g_{z}=\phi_{z}({\bf{v}}, {\bf{u}}) + \gamma({\bf{v}}, {\bf{u}}, \psi) + \eta_{z},
\label{eq:CH_Model_dB}
\end{equation}
where $\phi_{z}({\bf{v}}, {\bf{u}})$ is the path loss between the UAV and the user link, $\gamma({\bf{v}}, {\bf{u}}, \psi)$ stands for the antenna gain of the UAV with $\psi$ denoting the heading angle of the UAV with respect to the north pole, and $\eta_{z}$ is the shadowing component that is modeled as a Gaussian random variable with $\mathcal{N}(0,\sigma_{z}^{2})$. $z\in\left\{ \text{LoS},\text{NLoS}\right\}$ emphasizes the strong dependence of the propagation parameters on the \ac{los} or \ac{nlos} segments. The variance of the shadowing component $(\sigma_{z}^{2})$ is assumed to be known for both segments. Note that \eqref{eq:CH_Model_dB} represents the logarithm of the channel gain which is averaged over the small scale fading of unit variance.

Classically, the path loss $\phi_{z}({\bf{v}}, {\bf{u}})$ between  two radio nodes is modeled as \cite{ChenYanGes}
\begin{equation}
    \phi_{z}({\bf{v}}, {\bf{u}})\triangleq \phi_{z}(d) =\ss_{z}-10\,\alpha_{z}\log_{10}\left(d\right), \label{eq:classical_ch_gain}
\end{equation}
where $d = \|{\bf{v}} - {\bf{u}}\|_2$, $\alpha_{z}$ is the path loss exponent, and $\ss_{z}$ is the log of average path loss at the reference point $d=1\si{m}$. We assume that the users are equipped with an omnidirectional antenna. However, the antenna mounted on the UAV does not have a specific gain pattern and can have a very complex form depending on the design of the antenna and also the type of materials used in the UAV itself.


\section{Radio channel Learning and User Localization \label{sec:LearningLocalization}}

In this section, we propose a map-based algorithm to estimate the user locations from the channel gain measurements collected by the UAV. Let us denote an arbitrary set of measurements taken by the 
UAV during the mission by a sequence 
$\chi = \left\{{\bf{v}}_n, n\in [1,N]\right\}$.
From each of these 
locations, the UAV collects radio measurements form all $K$ users.
We denote the channel gain or RSS measurement (in dB scale) obtained from
the $k$-th user by the UAV in the $n$-th time step with $g_{n,k}$. Using the channel model in \eqref{eq:CH_Model_dB} we can write
\begin{equation}\label{eq:MeasurementModel}
g_{n,k} {=}
\begin{cases} 
    \phi_{\ac{los}}(d_{n, k}) + \gamma({\bf{v}}_n, {\bf{u}}_k, \psi_n) + \eta_{n,k,\ac{los}}       & \small{\text{if} \text{ LoS}}\\
    \phi_{\ac{nlos}}(d_{n, k}) + \gamma({\bf{v}}_n, {\bf{u}}_k, \psi_n) + \eta_{n,k,\ac{nlos}}       & \small{\text{if} \text{ NLoS}},
   \end{cases}
\end{equation} 
where $d_{n,k} = \| {\bf{v}}_n - {\bf{u}}_k \|_2$, and $\psi_n$ is the UAV heading angle at time step $n$. The function $\gamma (.)$ is the antenna gain which is unknown and it needs to be learned. The probability distribution of a single measurement in \eqref{eq:MeasurementModel} is modeled as
\begin{equation} \label{eq:mixdit}
    p(g_{n,k}) = (f_{n,k,\text{LoS}})^{w_{n,k}} (f_{n,k,\text{NLoS}})^{(1-w_{n,k})},
\end{equation}
where $\omega_{n,k} \in \{0,1\}$ is the classifier binary variable (yet unknown) indicating whether a measurement falls into the LoS or NLoS category, and $f_{n,k,z}$ has a Gaussian distribution with $\mathcal{N}(\phi_{z}(d_{n, k}) + \gamma({\bf{v}}_n, {\bf{u}}_k, \psi_n),\sigma_{z}^{2})$.

Assuming that collected measurements conditioned on the channel and user positions are independent and identically distributed (i.i.d) \cite{ChenYanGes}, using \eqref{eq:mixdit}, the negative log-likelihood of measurements leads to 
\noindent 
\begin{equation}
\begin{aligned}\label{eq:localization_liklihood}
\mathcal{L} &=  \log\left(\frac{\sigma_{\ac{los}}^2}{\sigma_{\ac{nlos}}^2}\right)\sum_{k=1}^{K}\sum_{n=1}^{N} \omega_{n,k}+ \\ &\sum_{k=1}^{K}\sum_{n=1}^{N}
  \frac{\omega_{n,k}}{\sigma_{\ac{los}}^2}\left |g_{n,k}{-}\phi_{\ac{los}}(d_{n, k}) - \gamma({\bf{v}}_n, {\bf{u}}_k, \psi_n)\right|^2 + \\
 & \sum_{k=1}^{K}\sum_{n=1}^{N} \frac{(1-\omega_{n,k})}{\sigma_{\ac{nlos}}^2}  \left | g_{n,k}{-}\phi_{\ac{nlos}}(d_{n, k}) - \gamma({\bf{v}}_n, {\bf{u}}_k, \psi_n)\right|^2.
\end{aligned}
\end{equation}
The estimate of the unknown channel parameters $\{\alpha_z, \ss_{z} \}$, $\bf{u}_k$, and $\gamma (.)$ can then be obtained by solving
\begin{subequations}
\begin{align}
 \begin{split}
           \min_{\substack{\omega_{n,k},\,{{\bf{u}}}_k\\ {\alpha_z, \ss_{z}, \gamma (.)}}
} & \quad \mathcal{L} 
 \end{split}\\
   \begin{split}
          \text{s.t.}&\quad \omega_{n,k} \in \{0,1\}, \forall n, \forall k.
 \end{split}
\end{align}\label{eq:Localization_Opt_Org}%
\end{subequations}
The binary variables $\omega_{n,k}$ in objective function \eqref{eq:localization_liklihood}, and the fact that $\gamma(.)$ is not explicitly known and is a function of user locations, make problem \eqref{eq:Localization_Opt_Org} challenging to solve since it is a joint classification, channel learning and user localization problem. To tackle this difficulty, we split \eqref{eq:Localization_Opt_Org} into two sub-problems of offline channel learning and online user localization . We also exploit the 3D map of the city for the measurements classification which will be elaborated next.

\subsection{Offline Radio Channel Learning} \label{sec:cahnnel_learning}
We aim to learn the radio channel using a set of offline radio measurements which are collected from users with known locations in advance. In this manner, we have a set of training data set to learn the radio channel. Since the characteristic of the radio channel is independent of the user location and only affected by the structure of the city (i.e. the blocking objects in the environment) and the UAV antenna pattern, therefore learning the radio channel from a set of offline training data set can provide a good approximation for the radio channel. We also exploit the 3D map of the city to perform the LoS/NLoS classification of the measurements, since for a user with a known location the classification variables $\omega_{n,k}$ can be directly inferred from a trivial geometry argument: for a given UAV position, the user is considered in LoS to the UAV if the straight line passing through the UAV's and the user position lies higher than any buildings in between. Moreover, we use a neural network with parameters ${{\boldsymbol{\theta}}}$ as an approximation of the UAV antenna gain $\gamma_{{{\boldsymbol{\theta}}}}(.)$. We call this channel model a \textit{hybrid} channel model, since it consists of a traditional path loss representation $\phi_{z}(.)$ along with a neural network approximating the UAV antenna gain. The main reason for choosing such a channel model lies in the fact that a rough estimation of the channel can be obtained using the classical path loss model \eqref{eq:classical_ch_gain}, while all the uncertainties which can not be captured by the path loss function are then modeled using a neural network.

Now having classified the measurement and using the hybrid channel model, problem \eqref{eq:Localization_Opt_Org} just by considering the offline training data set (with known user locations) can be rewritten as follows
\begin{equation}
\begin{aligned}
             \min_{\substack{{\alpha_z, \ss_{z}, {\boldsymbol{\theta}}}}
} & \quad \mathcal{L},
\end{aligned} \label{eq:cahnnel_learning}
\end{equation}
where $\theta$ is the parameters of the neural network for estimating of the antenna gain. Solving this problem is still challenging since the UAV antenna gain is the same for both LoS and NLoS measurements. To alleviate this burden, we split up our problem into two phases. In the first phase, we only find the path loss parameters by solving the following optimization problem 
\begin{equation}
\begin{aligned}
             {\alpha_z^*, \ss_{z}^*} := \argmin_{\substack{{\alpha_z, \ss_{z}}}
} & \quad \Bar{\mathcal{L}},
\end{aligned} \label{eq:cahnnel_learning_p1}
\end{equation}
where 
\begin{equation}
\begin{aligned}\label{eq:L_p1}
\Bar{\mathcal{L}} =&  \sum_{k=1}^{K}\sum_{n=1}^{N}
  \frac{\omega_{n,k}}{\sigma_{\ac{los}}^2}\left |g_{n,k}{-}\phi_{\ac{los}}(d_{n, k}) \right|^2 + \\
 & \sum_{k=1}^{K}\sum_{n=1}^{N} \frac{(1-\omega_{n,k})}{\sigma_{\ac{nlos}}^2}  \left | g_{n,k}{-}\phi_{\ac{nlos}}(d_{n, k})\right|^2.
\end{aligned}
\end{equation}
In \eqref{eq:L_p1} the effect of the UAV antenna gain is ignored which allows us to find the closest estimate to the measurements using the path loss model. The parameters obtained by solving \eqref{eq:cahnnel_learning_p1} are denoted as ${\alpha_z^*, \ss_{z}^*}$.


In the second phase, the path loss parameters are fixed to ${\alpha_z^*, \ss_{z}^*}$ and the UAV antenna gain parameters are obtained as follows
\begin{equation}
\begin{aligned}
            {\boldsymbol{\theta}}^* := \argmin_{\substack{{{\boldsymbol{\theta}}}}
} & \quad \mathcal{L}|_{\alpha_z^*, \ss_{z}^*}.
\end{aligned} \label{eq:cahnnel_learning_p2}
\end{equation}
Note that, both problems \eqref{eq:cahnnel_learning_p1}, \eqref{eq:cahnnel_learning_p2} can be solved using standard optimization frameworks (i.e. any gradient-based optimizer).

\subsection{User Localization} \label{sec:localization}
Having learned the radio channel, we continue to localize the unknown users in the online data set. The optimization problem \eqref{eq:Localization_Opt_Org} by utilizing the learned radio channel can be reformulated as follows:
\begin{subequations}
\begin{align}
  \begin{split}
           \min_{\substack{\omega_{n,k},\,{{\bf{u}}}_k}
} & \quad \mathcal{L}^* 
 \end{split}\\
   \begin{split}
          \text{s.t.}&\quad \omega_{n,k} \in \{0,1\}, \forall k , \forall n,
 \end{split}
\end{align}\label{eq:Localization_unkown_users}%
\end{subequations}
where $\mathcal{L}^* $ is obtained by substituting the channel model with learned parameters ${\alpha_z^*, \ss_{z}^*}, {\boldsymbol{\theta}}^*$ in \eqref{eq:localization_liklihood}. It is hard to find a closed form and analytical solution to problem \eqref{eq:Localization_unkown_users} due to the binary random variables $\omega_{n,k}$, and the non-linear and non-convex objective function $\mathcal{L}^*$. We employ the  particle swarm optimization (PSO) technique to solve this problem since PSO is suitable for solving various non-convex and non-linear optimization problems. More specifically, PSO is a population-based optimization technique that tries to find a solution to an optimization problem by iteratively trying to improve a candidate solution with regard to a given measure of quality (or objective function). The algorithm is initialized with a population of random solutions, called particles, and a search for the optimal solution is performed by iteratively updating each particle's velocity and position based on a simple mathematical formula (for more details on PSO see \cite{KenEbe}). As will be clear later, the PSO algorithm is enhanced to exploit the side information stemming from the 3D map of the environment which improves the performance of user localization and reduce the complexity of solving $\eqref{eq:Localization_unkown_users}$, since the binary variables $\omega_{n,k}$ can be obtained directly from the 3D map \cite{esrafilian20203d}.

For ease of exposition, we first solve $\eqref{eq:Localization_unkown_users}$ by assuming only one unknown user. Then we will generalize our proposed solution to the multi-user case. To apply the PSO algorithm, we define each particle to have the following form
\begin{equation}
    {\bf{c}}_j = [ x_j, y_j]^{\text{T}} \in \mathbb{R}^2, j \in [1, C],
\end{equation}
where $C$ is the number of particles and each particle is an instance of the possible user location in the city. Therefore, by treating each particle as a potential  candidate for the user location, the negative log-likelihood $\eqref{eq:localization_liklihood}$ for a given particle and learned parameters ${\alpha_z^*, \ss_{z}^*}, {\boldsymbol{\theta}}^*$ can be rewritten as follows

\begin{equation}
\begin{small}
\begin{aligned} 
\mathcal{L}^*(&{\bf{c}}_{j}^{(i)}) = \log\left(\frac{\sigma^2_{\ac{los}}}{\sigma^2_{\ac{nlos}}}\right) \left |\mathcal{M}_{\ac{los},1,j}\right| + \\ & \small{\sum_{z\in \{\ac{los},\ac{nlos}\}} \, \sum_{n\in \mathcal{M}_{z,1,j}}} \frac{1}{\sigma^{2}_{z}}
 \left |g_{n,1}{-}\phi_{z}(d_{n, j}) - \gamma_{\theta^*}({\bf{v}}_n, {\bf{c}}_{j}^{(i)}, \psi_n)\right|^2  ,\label{eq:SSE_PSO_single_Ue}
\end{aligned}
\end{small}
\end{equation}

\noindent where ${\bf{c}}_{j}^{(i)}$ is the $j$-th particle at the $i$-th iteration of the PSO algorithm, $d_{n, j} = \|{\bf{v}}_n - {\bf{c}}_{j}^{(i)}\|_2$, and $\mathcal{M}_{z,1,j}$ is a set of time indices of measurements collected from user 1 which are in segment $z$ by assuming that the location of user 1 is the same as particle $j$. To form $\mathcal{M}_{z,1,j}$, a 3D map of the city is utilized. For example, measurement $g_{n,1}$ is considered LoS, if the straight line passing through ${\bf{c}}_{j}^{(i)}$ and the drone location ${\bf{v}}_n$ lies higher than any buildings in between. Therefore, the best particle minimizing \eqref{eq:SSE_PSO_single_Ue} can be obtained from solving the following optimization
\noindent 
\begin{equation}
    j^* := \arg \min_{j\in [1,C]} {\mathcal{L}^{*}({\bf{c}}_{j}^{(i)})}, \label{eq:best_particle}
\end{equation}
where $j^* $ is the index of the best particle which minimizes the objective function in \eqref{eq:best_particle}. In the next iteration of the PSO algorithm, the position and the velocity of particles are updated and the algorithm repeats for $\tau$ iterations. The best particle position in the last iteration is considered as the estimate of the user location.

Note that for the multi-user case, without loss of optimality, the problem can be transformed into multiple single-user localization problems, and then each problem can be solved individually. This stems from the fact that the radio channel is learned beforehand and is assumed to have the same characteristics for all the UAV-user links (the radio channel parameters and the UAV antenna pattern are independent of user locations).

\begin{figure}[t]
\centering
\subfloat[]{
  \includegraphics[clip,width=0.85\columnwidth]{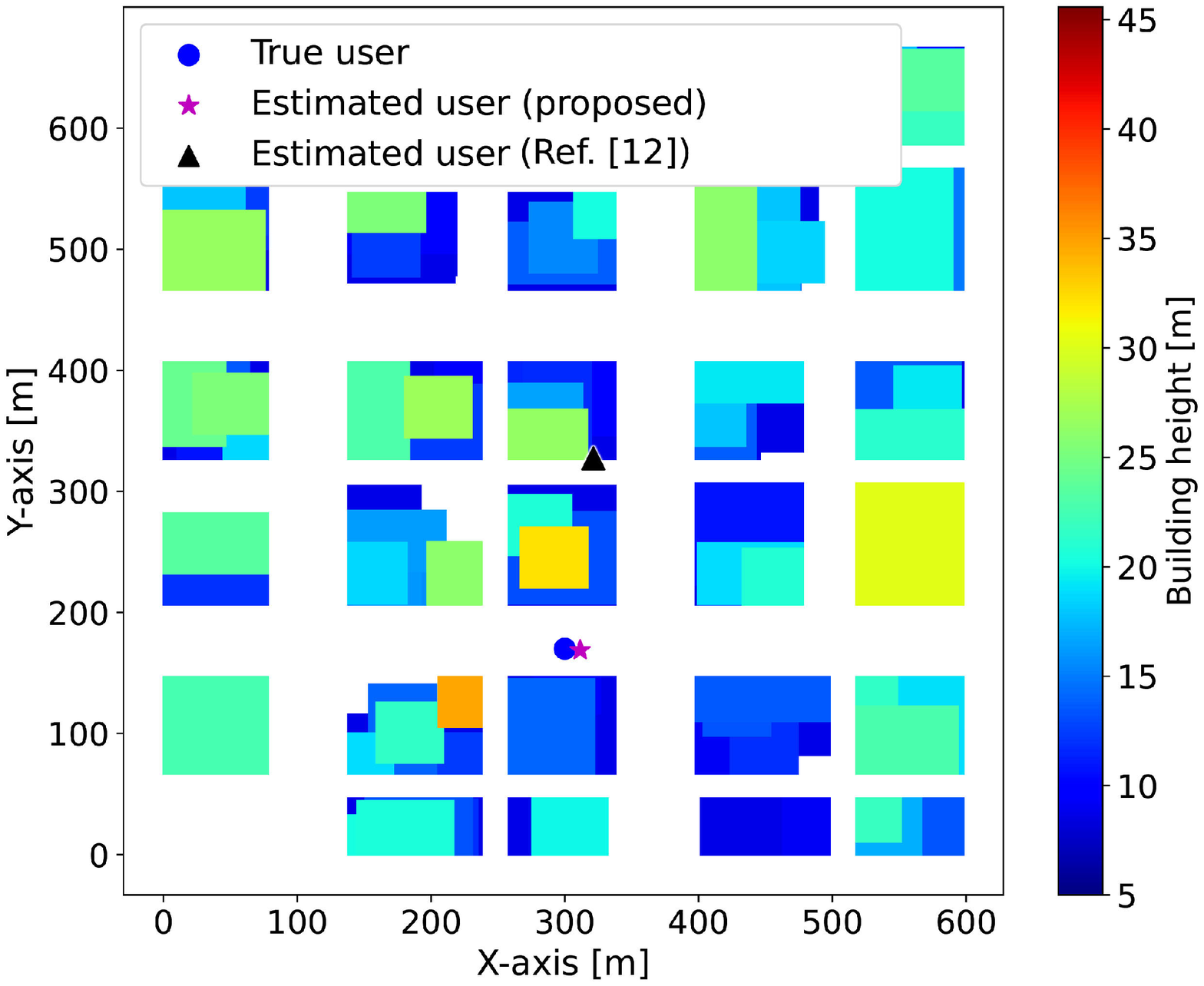}
}
\newline
\centering
\subfloat[]{
  \includegraphics[clip,width=0.85\columnwidth]{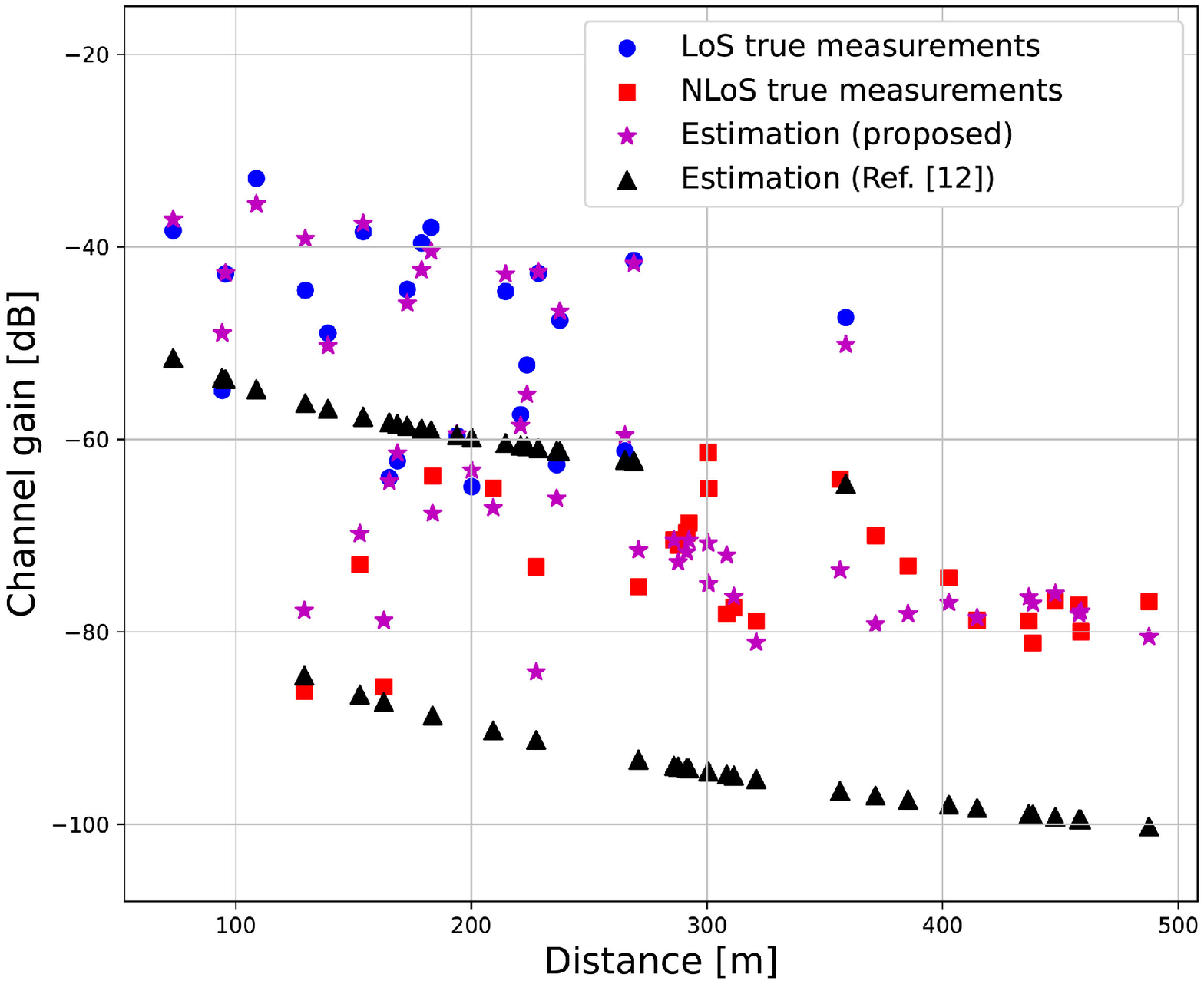}
}
\caption{(a) Localization performance by utilizing the proposed hybrid channel model and the conventional channel model, (b) Corresponding test radio measurements collected from the same user and the estimation using different algorithms. \label{fig:hybrid_localziation}}
\end{figure}

\section{Numerical Results}\label{sec:simulations}
We consider a dense urban city neighborhood comprising buildings and streets as shown in Fig. \ref{fig:hybrid_localziation}-a. The height of the buildings is Rayleigh distributed in the range of 5 to \SI{40}{m} \cite{Ref26_HourKandeepJamail}. The true propagation parameters are chosen as  $\alpha_{\ac{los}}=2.2,\,\alpha_{\ac{nlos}}=3.2,\,\ss_{\ac{los}}=-32\,\text{dB},\,\ss_{\ac{nlos}}=-35\,\text{dB}$
according to an urban micro scenario \cite{3GPP}. The variances
of the shadowing components in \ac{los} and \ac{nlos} scenarios are
$\sigma_{\ac{los}}^{2}=2\,\text{dB}$, and $\sigma_{\ac{nlos}}^{2}=5\,\text{dB}$, respectively. The following UAV antenna gain is considered to conduct the simulation
\begin{equation}
    \gamma({\bf{v}}_n, {\bf{u}}_k, \psi_n) = 15  \left(| cos(\rho_{n, k})| + 2  |sin(\varphi_{n, k} + \psi_n)|\right),
\end{equation}
where $\rho_{n, k}, \varphi_{n, k}$ are , respectively, the elevation and azimuth angles between the UAV and the user, and $\psi_n$ is the heading angle of the UAV.
The hybrid channel model is trained in the same city as shown in Fig. \ref{fig:hybrid_localziation}-a by collecting radio measurements from $K=10$ different random users and over $N=200$ individual random UAV locations. To train the hybrid model, we first need to learn the pathloss parameters. To do so, we use the training data set $\mathcal{D}^{tr}_{pl} = \{ (d_{n, k}, g_{n, k}), \forall n, k \}$. The path loss parameters can then be obtained by solving \eqref{eq:cahnnel_learning_p1}.
Now we continue to learn the UAV antenna gain. To estimate the UAV antenna gain, a neural network with four hidden layers is used where the first and the second layers have $60$ neurons with the $tanh$ activation function, and the third and the fourth layers with $40$ neurons and the $relu$ activation function. To train this network the training data set $\mathcal{D}^{tr}_{ag} = \{ (d_{n, k}, {\bf{x}}_{n, k}, g_{n, k}), \forall n, k \}$ is used where ${\bf{x}}_{n, k}$ is the input vector to the neural network and is defined as follows
\begin{equation}
    {\bf{x}}_{n, k} = [\frac{x[n] - x_k}{\|{\bf{v}}_n - {\bf{u}}_k\|_2}, \frac{y[n] - y_k}{\|{\bf{v}}_n - {\bf{u}}_k\|_2}, \frac{z[n]}{\|{\bf{v}}_n - {\bf{u}}_k\|_2}, \psi_n]^{\text{T}}.
\end{equation}
The parameters of the neural network $(\theta)$ can be obtained by solving the optimization problem \eqref{eq:cahnnel_learning_p2} where $\mathcal{L}$ is defined as follows

\begin{equation}
\begin{aligned}
\mathcal{L} &=  \log\left(\frac{\sigma_{\ac{los}}^2}{\sigma_{\ac{nlos}}^2}\right)\sum_{k=1}^{K}\sum_{n=1}^{N} \omega_{n,k}+ \\ &\sum_{k=1}^{K}\sum_{n=1}^{N}
  \frac{\omega_{n,k}}{\sigma_{\ac{los}}^2}\left |g_{n,k}{-}\phi_{\ac{los}}(d_{n, k}) - \gamma_{\theta}({\bf{x}}_{n, k})\right|^2 + \\
 & \sum_{k=1}^{K}\sum_{n=1}^{N} \frac{(1-\omega_{n,k})}{\sigma_{\ac{nlos}}^2}  \left | g_{n,k}{-}\phi_{\ac{nlos}}(d_{n, k}) - \gamma_{\theta}({\bf{x}}_{n, k})\right|^2.
\end{aligned}
\end{equation}

Having trained the hybrid channel model, the estimate of each measurement in accordance with \eqref{eq:MeasurementModel} is given by:
\begin{equation}\label{eq:mixed_ch_model_with_input}
\hat{g}_{n,k} {=}
\begin{cases} 
    \phi_{\ac{los}}(d_{n, k}) + \gamma_{\theta}({\bf{x}}_{n, k})      & \small{\text{if} \text{ LoS}}\\
    \phi_{\ac{nlos}}(d_{n, k}) + \gamma_{\theta}({\bf{x}}_{n, k})        & \small{\text{if} \text{ NLoS}}.
   \end{cases}
\end{equation}

\begin{figure}[t]
\begin{centering}
\includegraphics[width=0.8\columnwidth]{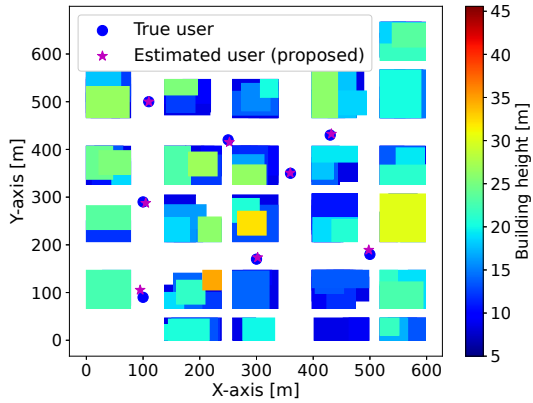}
\par\end{centering}
\caption{The performance of proposed localization algorithm for multi-user case.  
\label{fig:multiUser_localization}}
\vspace{-6pt}
\end{figure}

In Fig. \ref{fig:hybrid_localziation}-a, the result of the user localization after the training phase is shown. We also compared the performance of the proposed algorithm with \cite{esrafilian20203d} which uses the conventional channel model consisting of the path loss model without considering the effect of the UAV antenna gain. Moreover, in Fig. \ref{fig:hybrid_localziation}-b the results of the channel model estimation is shown for different algorithms. It is clear that by using the hybrid channel model we can obtain a better estimation of the channel which results in a more precise user localization. Moreover, the user location estimated using the proposed algorithm for the multi-user scenario is illustrated in Fig. \ref{fig:multiUser_localization} and confirmed to be very close to the true user positions.

\begin{figure}[t]
\begin{centering}
\includegraphics[width=0.8\columnwidth]{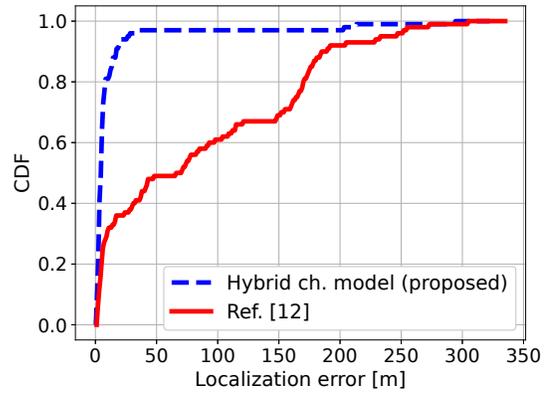}
\par\end{centering}
\caption{The CDF of user localization error for different algorithms.  
\label{fig:cdf_localization}}
\vspace{-6pt}
\end{figure}

In Fig. \ref{fig:cdf_localization}, the cumulative distribution function (CDF) of user localization error of our proposed algorithm with comparison to \cite{esrafilian20203d} over Monte-Carlo simulations is shown. We can see that the localization accuracy is considerably improved by using the hybrid channel model.   

\section{Experimental Results}\label{sec:experiment}
We have also validated the performance of the proposed algorithm through real-world experimentation. We established a Wi-Fi mesh network comprising sets of outdoor ground nodes and a UAV node. all the nodes are equipped with a MicroTick Wi-Fi card, which is configured on channel 48 in the 5 GHz band, with two omnidirectional vertically polarized dipole antennas. Prior to applying the localization algorithm, we learned the wireless channel by training the hybrid channel model over the
training measurements collected from different ground users in the environment where the experiment is conducted.

\begin{figure}[t]
\centering
\subfloat[]{
  \includegraphics[clip,width=0.49\columnwidth]{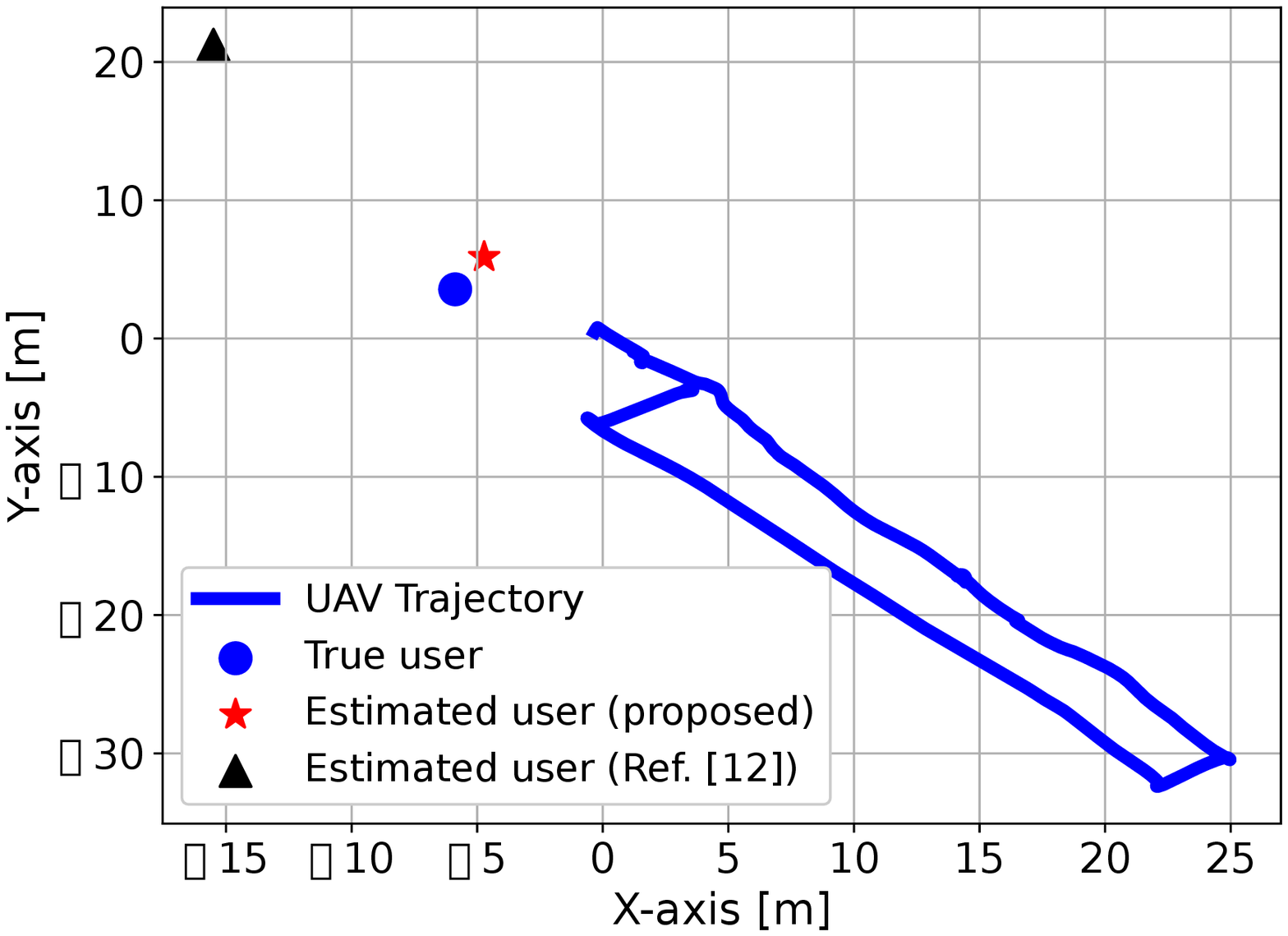}
  \includegraphics[clip,width=0.48\columnwidth]{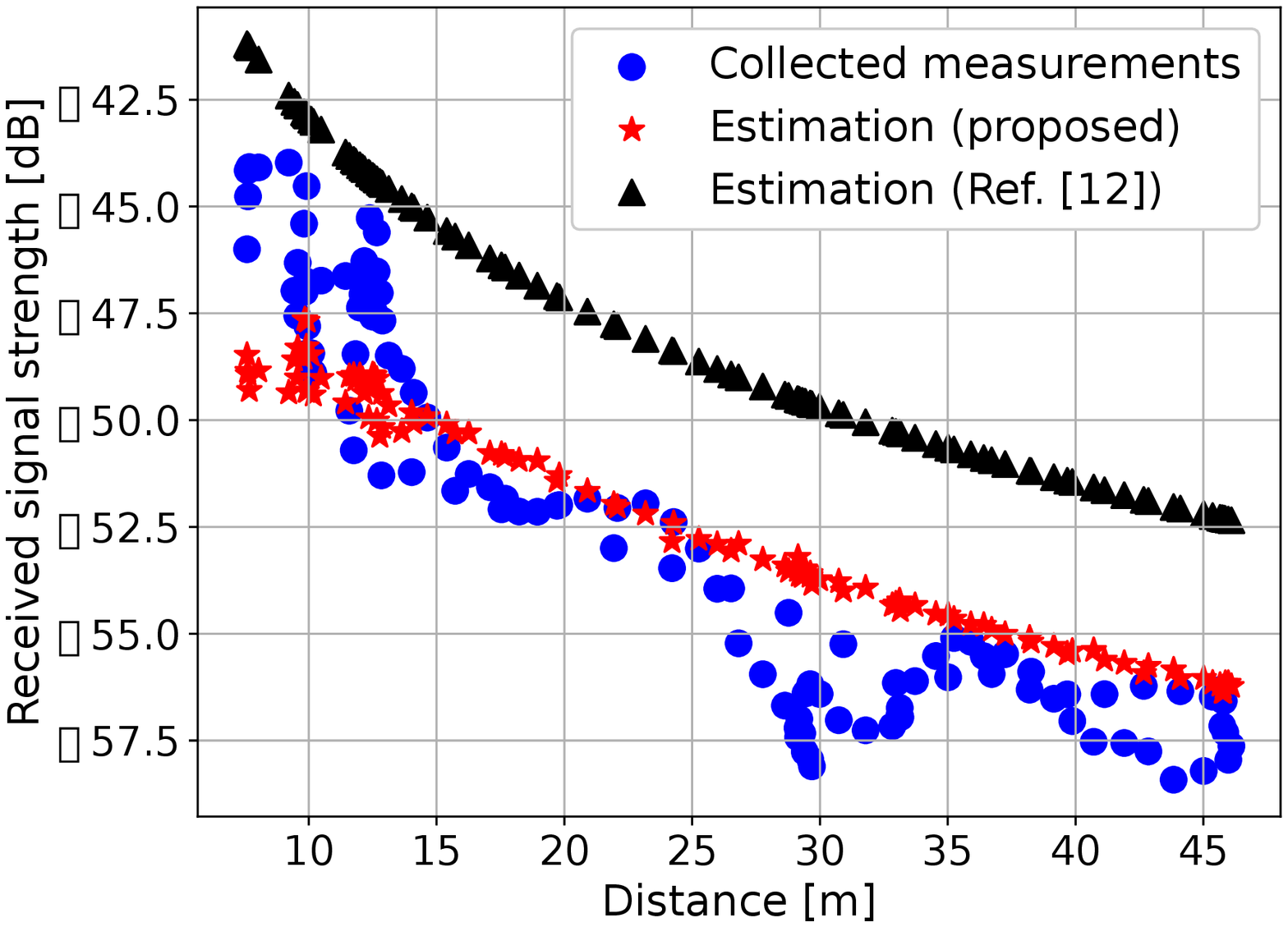}
}
\newline
  \subfloat[]{
  \includegraphics[clip,width=0.49\columnwidth]{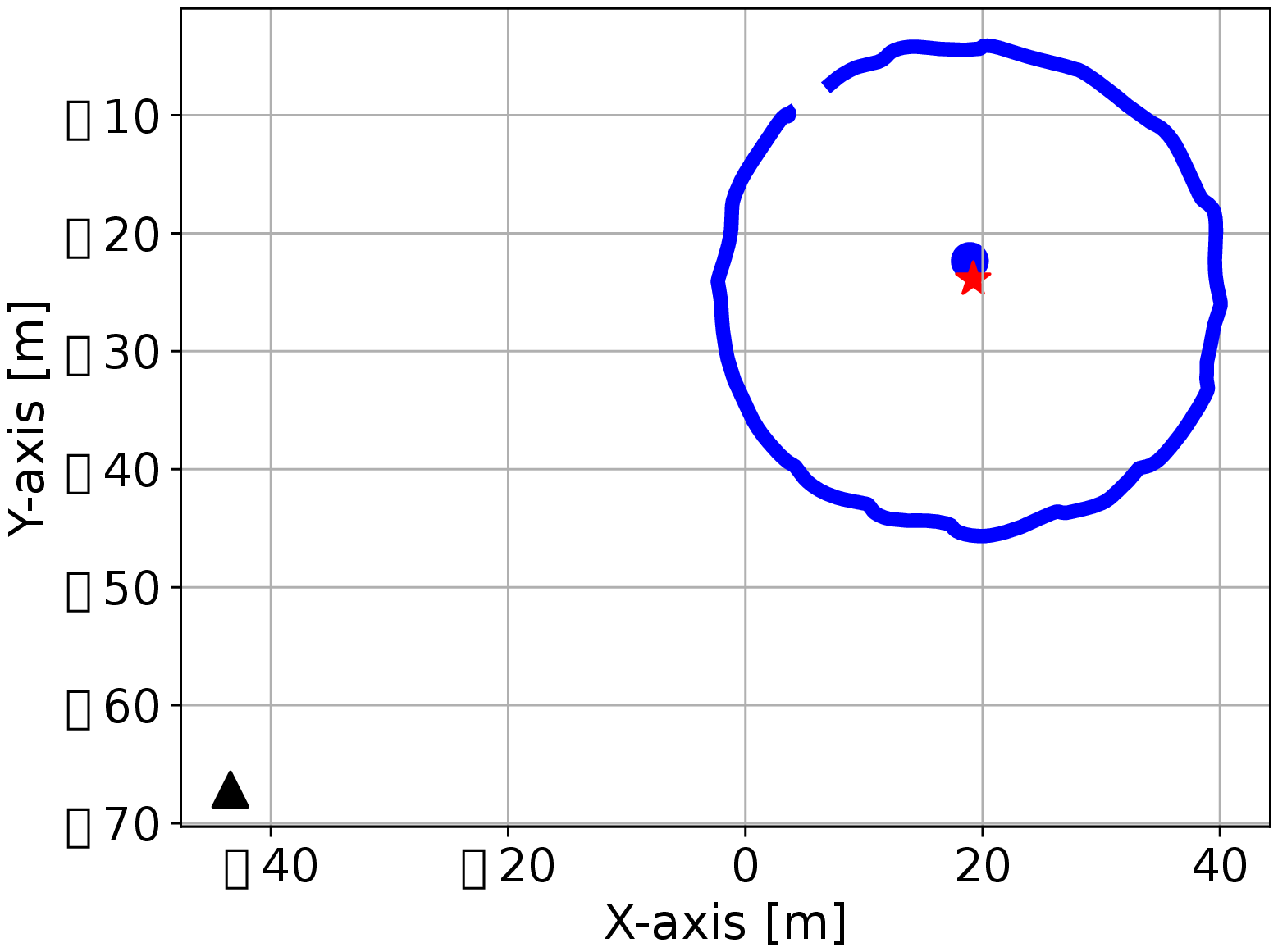}
  \includegraphics[clip,width=0.48\columnwidth]{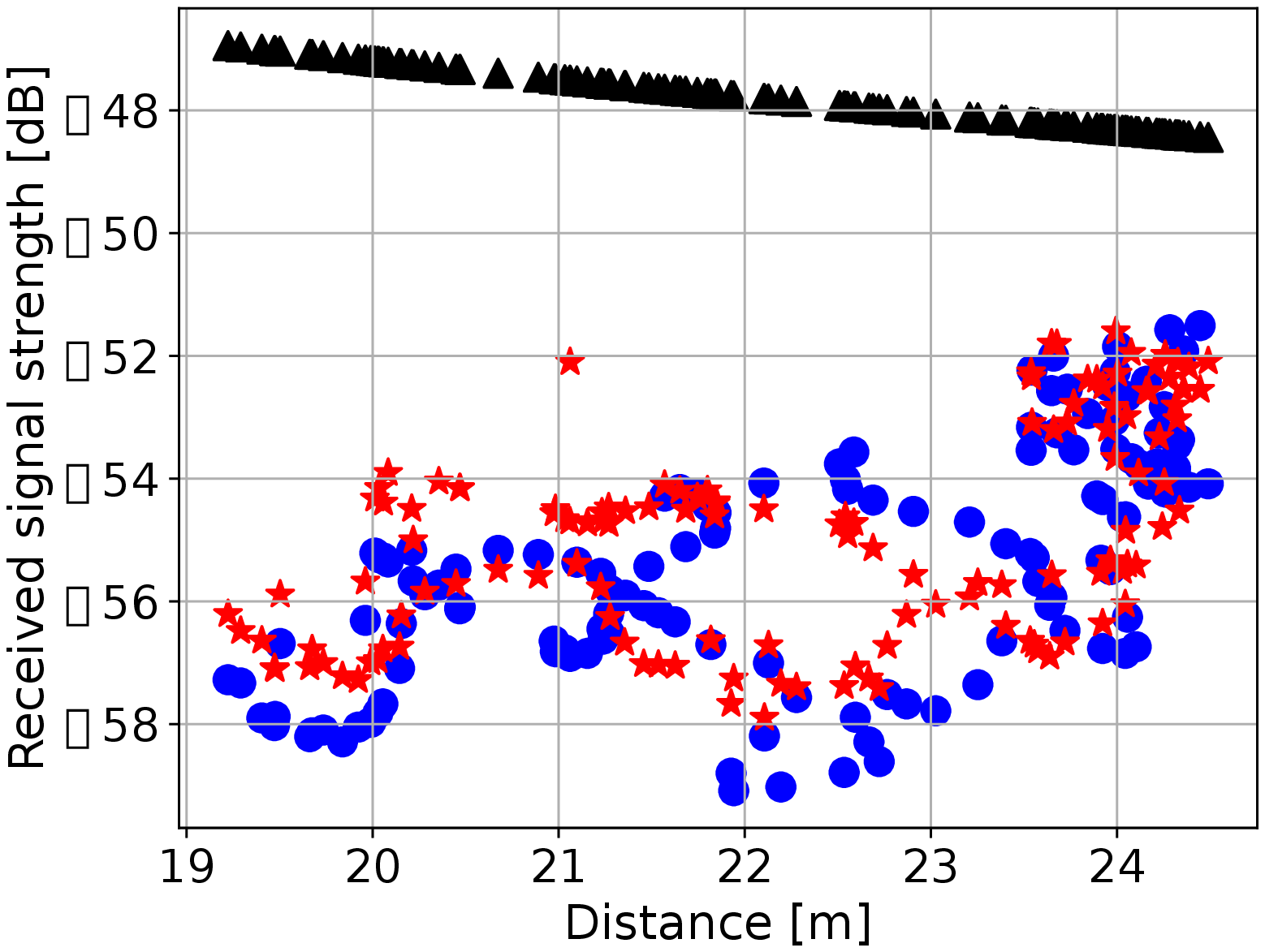}
}
\caption{Left: top view of the UAV trajectory, true user and estimated user location. Right: corresponding test measurements collected from the user and the channel estimate using different models. \label{fig:experiment_localization}}
\end{figure}

Having learned the channel model, different test user locations are chosen to be localized. In Fig. \ref{fig:experiment_localization}, the top view of the UAV trajectory taken to collect the test data, the performance of the localization, and the estimate of the channel using our proposed method as well as the method in \cite{esrafilian20203d} for different scenarios are shown. In both trajectories, the heading of the UAV is set to a fixed angle facing towards the south over the course of the trajectory. It is worth noting that when the relative angle between the UAV and the ground node changes drastically, Fig. \ref{fig:experiment_localization}-b, the algorithm in [12] which uses the convectional channel model fails to localize the users. This stems from the fact that the UAV antenna pattern is changed considerably and is no longer symmetric and omnidirectional due to the proximity to all the components on the UAV (i.e. the body frame, propellers, motors, etc.) which makes it difficult to be precisely modeled by conventional channel models.

Moreover, in \cite{VideoClip}, a video recording
of the experiment in EURECOM campus is captured, illustrating the localization of two ground nodes while flying the UAV in the environment. As the UAV collects more measurements from the ground nodes, the estimate of the user location becomes more accurate.

\section{Conclusion}\label{sec:conclusion}
We have studied the problem of user localization using the RSS measurements collected by a UAV. To do so, we first proposed a hybrid channel model which aims to accurately learn the path loss parameters as well as the UAV antenna pattern. A PSO technique then was employed by exploiting the learned hybrid channel model and leveraging the 3D map of the environment to accurately localize the ground users. The performance of the developed algorithm was evaluated through simulations and also real-world experiments.

\section{Acknowledgments}
This work was funded via the HUAWEI
France supported Chair on Future Wireless Networks at EURECOM.

\bibliographystyle{IEEEtran}
\bibliography{literature.bib}

\end{document}